%% file: paper.tex
\documentclass[a4paper,11pt]{article}
\usepackage{amsmath,amssymb,amsthm,amsfonts}    % AMS math packages
\usepackage{mathrsfs,bm}                        % Bold math symbols
\usepackage{dsfont}
\usepackage{eurosym}                            % Euro symbol
\usepackage{xcolor}                             % Color text management
\usepackage{url,hyperref}                       % Urls and hypertexts (for online files) management
\usepackage{float}                              % Floating objects management
\usepackage[small,bf]{caption}                  % Advanced caption management
\usepackage{graphicx}
\usepackage{epstopdf}                           % Figures management
\usepackage{subfigure}
\usepackage{psfrag}                             % Allows to overlay EPS figures with arbitrary LaTeX constructions
\usepackage{epsfig}                             % Encapsulated postscript management
\graphicspath{{./Figures/}}                     % Figures path.
\usepackage{booktabs,multirow,longtable}        % Advanced table management
\usepackage{tabularx}        % Advanced table management
\usepackage[toc,page]{appendix}
\usepackage[english]{babel}
\usepackage{mathtools}
\usepackage{makecell}
\usepackage[margin=0.5in]{geometry}
\usepackage{booktabs}
\usepackage{longtable}
\usepackage[nottoc,notlot,notlof]{tocbibind}
\usepackage{comment}
\usepackage{adjustbox}
\makeatletter \@addtoreset{equation}{section} \makeatother

\begin{document}
\include{Newcommands}	% File with many useful LaTeX commands

\title{\textbf{Learning Bermudans}}

\author{
    Riccardo Aiolfi\footnote{Dipartimento di Fisica \QuoteDouble{Aldo Pontremoli}, Università degli Studi di Milano, via Giovanni Celoria 16, 20133 Milano, Italy, \texttt{riccardo.aiolfi@studenti.unimi.it}}, Nicola Moreni\footnote{Intesa Sanpaolo, Financial Engineering, IR\&Credit Models, \texttt{nicola.moreni@intesasanpaolo.com}},	Marco Bianchetti\footnote{Intesa Sanpaolo, Market, Financial and C\&IB Risks, and University of Bologna, Department of Statistical Sciences \QuoteDouble{Paolo Fortunati}, \texttt{marco.bianchetti@unibo.it}}, Marco Scaringi\footnote{Intesa Sanpaolo, Market, Financial and C\&IB Risks, \texttt{marco.scaringi@intesasanpaolo.com}}, Filippo Fogliani\footnote{Intesa Sanpaolo, Market, Financial and C\&IB Risks, \texttt{filippo.fogliani@intesasanpaolo.com}}
}

\date{\today}
\maketitle

\begin{abstract}
American and Bermudan-type financial instruments are often priced with specific Monte Carlo techniques whose efficiency critically depends on the effective dimensionality of the problem and the available computational power. In our work we focus on Bermudan Swaptions, well-known interest rate derivatives embedded in callable debt instruments or traded in the OTC market for hedging or speculation purposes, and we adopt an original pricing approach based on Supervised Learning (SL) algorithms. In particular, we link the price of a Bermudan Swaption to its natural hedges, i.e. the underlying European Swaptions, and other sound financial quantities through SL non-parametric regressions. We test different algorithms, from linear models to decision tree-based models and Artificial Neural Networks (ANN), analyzing their predictive performances. All the SL algorithms result to be reliable and fast, allowing to overcome the computational bottleneck of standard Monte Carlo simulations; the best performing algorithms for our problem result to be Ridge, ANN and Gradient Boosted Regression Tree. Moreover, using feature importance techniques, we are able to rank the most important driving factors of a Bermudan Swaption price, confirming that the value of the maximum underlying European Swaption is the prevailing feature. 
\end{abstract}

\newpage
\tableofcontents

\vspace{2cm} 

\vspace{0.5cm} 

\vspace{0.5cm} 
\noindent \textbf{Acknowledgements}: The authors acknowledge fruitful discussions with Davide Emilio Galli and thank him for his constructive comments.

\vspace{0.5cm} 
\noindent \textbf{Disclaimer}: The views expressed here are those of the authors and do not represent the opinions of their employers. They are not responsible for any use that may be made of these contents.

\newpage

%%%%%%%%%%%%%%%%%%%%%%%%%%%%%%%%%%%%%%%%%%%

\section{Introduction}
\label{Sec:Intro}

In the very last years, a variety of Machine Learning (ML) techniques has been successfully employed in quantitative finance. An approach that uses ML algorithms can be of great help to get more insight and information about very complex systems as well as being able to provide more efficient solutions than current ones. In this paper we will deal exclusively with the Supervised Learning (SL) branch of ML. SL algorithms are those that automate decision-making processes by generalizing from known examples. Leveraging on these techniques in this paper we face the optimal stopping time problem. In the context of complex systems, an optimal stopping time can be generally thought of as a strategy about when to perform an action in order to maximize an expected reward or minimize an expected cost. Within the field of quantitative finance, optimal stopping time problems arise for instance when pricing a class of products known as American or Bermudan-type options; in these contracts, the buyer has the right to enter a financial transaction at the time which is considered as the best one. In this case, the optimal stopping time is the one that maximizes the gain for the investor.

In theory, optimal stopping problems with a great but finite number of stopping opportunities can be solved analytically or semi-analytically but, being very complex, they have been in practice studied extensively using the methodology of dynamic programming coupled to traditional Monte Carlo simulations. Several algorithms can be found in literature \cite{Gla03}. Due to the curse of dimensionality, these problems are typically tied to the efficiency of the Monte Carlo methods and above all to the computational power available. Recently, authors have proposed the use of ML algorithms to solve or speed up optimal stopping time problems. For instance, \cite{Bec20} after setting up a quite general framework show some practical applications, \cite{Gas20}, \cite{BecCher20} and \cite{Lap20} directly focus on the American/Bermudan option pricing problem through deep learning and finally \cite{Gou19} also suggest variance reduction techniques. All of these approaches employ a subset of ML algorithms belonging to the field of Artificial Neural Networks (ANNs) to solve dynamic programming or to approximate the optimal exercise boundary, but they are still based on traditional Monte Carlo numerical simulations which can be burdensome for high dimensional systems. 

The present work aims to contribute to this area of research with an original approach based on a completely different point of view. In particular, we try to link the price of an interest rate swap Bermudan option (Bermudan Swaption), well-known interest rate derivatives embedded in callable debt instruments or traded in the OTC market for hedging or speculation purposes, to financial quantities that are available to market participants and are its main drivers, the so-called natural hedges. Focusing on different SL algorithms we performed non-parametric regressions to get estimators for option prices and we carried out a comparative analysis between them to find the best one. We believe that our approach could successfully address these problems overcoming the computational bottleneck of standard Monte Carlo numerical simulations. Furthermore, our techniques could help us to understand the most important driving factors behind the market price. To explore many market scenarios, regarding the variance and covariance of relevant financial quantities and thus make our work as general as possible, we have generated a synthetic coherent price dataset through numerical simulations based on the interest short model developed by Hull and White in \cite{HulWhi1994} and subsequent papers. Our synthetic target market prices were obtained through a dynamic programming algorithm known as Least Square Monte Carlo (LSMC) \cite{LonSch1998}.

This paper is made of five chapters; to make this works as self-consistent as possible, in the first chapter we give a short compendium of the tools implemented for this work included the Hull-White One Factor model, the LSMC and the SL algorithms considered. The second chapter is completely dedicated to the creation of the dataset and its inspection while in the third chapter we present the numerical results obtained including the final comparative analysis between the algorithms' performance and the feature importance. The last chapter is devoted to the conclusion of our work and possible future perspectives.

\section{Theoretical Setting}
\label{Sec:Theory}

In this chapter we give a short compendium of the tools implemented for this work. The first two sections briefly deal with the financial topics at the centre of our research, while the last section introduces the SL and the algorithms considered.

\subsection{Bermudan Swaptions}
\label{Subsec:bermudan_swaption}

Swaptions are interest rate derivatives on an Interest Rate Swap (IRS) typically traded by large corporations, banks, financial institutions, and hedge funds. There are two main versions of Swaptions, a payer and a receiver. A payer Swaption is an option giving the right (and no obligation) to enter a payer IRS at a given future time, the Swaption maturity; in other words, the buyer has the right to become the fixed rate payer in an IRS, which length is called the tenor of the Swaption. Instead in the receiver version, the buyer has the right to become the receiver of the fixed leg. There are two standard market payoffs, that differ in the settlement convention: physical or cash. We will focus only on the first type, i.e those once exercised, are transformed into the underlying swap. In general, three main styles define the exercise of  derivative instruments and therefore also of a Swaption: European, Bermudan and American. In this work we will focus only on Bermudan Swaptions, i.e. exotic interest rate derivatives that allows the buyer to enter, at multiple exercise dates $\left\{ T_{1},\dots,T_{N} \right\}$ into a swap starting at time $T_{0} \geq T_{i}$ and maturing at $T_{m}$. Notice that European Swaptions can be seen as Bermudan with a single exercise date and in turn, the American type can be seen as the extension to the continuum of the Bermudan. Bermudan Swaptions are an example of derivative embedding early exercise and no market quotations or broker pages are available, in fact they cannot be priced analytically, since their value depends, at each exercise date, on the choice of the option holder whether it is more convenient to exercise it (retrieving the payoff) or to continue with the contract (continuation value).

\subsection{Hull White One-Factor Model and Least Square Monte Carlo}
\label{Subsec:hull_lsmc}

To analyze and price the type of instruments described in the previous section we implemented two tools: the Hull-White One-Factor Model (G1++) \cite{HulWhi1994} and the Least Square Monte Carlo (LSMC) \cite{LonSch1998}. 

The Hull-White One-Factor Model, also known as G1++, is a specific case of the Ornstein-Uhlenbeck process characterized by a single stochastic factor. It is one of the major exogenous short rate models which is nowadays often used for pricing and risk management purposes and specifically we used it for the simulation of the underlying stochastic dynamics and hence for the evolution of the interest rate curve.  This model is analytically tractable, in fact there are closed pricing formulas for some instruments, e.g. European Swaptions; this feature is decisive for us as European Swaptions represent the natural hedges of Bermudan Swaptions and they have a fundamental role in the pricing of these products \cite{Hag02}. Our aim is to probe different market scenarios, but since in recent years the rates and their correlations have always been low, these historical data do not allow us to have enough wealth in the dataset. For this reason, we have exploited the two G1++ parameters, i.e. speed of mean reversion $(a)$ and volatility $(\sigma)$, to create many different market scenarios that differ in the global level of variances and covariances of the relevant stochastic processes, in order to increase the variability of our dataset, avoiding any type of calibration.

Instead, the Least Square Monte Carlo (LSMC) is one of the most widely used dynamic programming tools for the pricing of American-type option. It is one of the methods proposed to reduce the complexity of American option pricing avoiding nested Monte Carlo; it is a regression-based method that use some specific function (basis function) to approximate the continuation values in the underlying optimal stopping time problem. The success of this type of method, as well as depending on the computational power available, strongly relies on the choice of the basis functions and their number, making it is still tied to the efficiency of the Monte Carlo simulation.

\subsection{Supervised Learning algorithms}
\label{Subsec:sl_algorithms}

The implementation of the tools explained in previous sections allowed us, starting from real market data (Appendix \ref{App:market}) to obtain a synthetic price dataset used to train the supervised algorithms. To find out which supervised algorithm is best suited to our problem, we analysed a very heterogeneous set of models. Generally, since the problems faced with these techniques involve inferences on complex systems, it is a common choice to select several candidate models to which their predictive performance must be compared. Below we present the list of algorithms used in this work. For a more in-depth discussion on their main characteristics, strengths and weaknesses we refer to Appendix \ref{App:SL}, while for all mathematical details we refer to \cite{has01,ger17}.
\begin{itemize}
    \item k-Nearest Neighbour (k-NN);
    \item Linear Models;
    \item Support Vector Machine (SVM);
    \item Tree-based algorithms;
    \begin{itemize}
        \item Random Forest (RF);
        \item Gradient Boosted Regression Tree (GBRT).
    \end{itemize}
    \item Artificial Neural Networks (ANN).
\end{itemize}
Although these algorithms are all different from each other, in order to define their predictive capabilities and to be able to compare them we have adopted a similar approach for all. It can be divided into 3 steps:
\begin{enumerate}

\item The first focuses on the modelling of the input data in such a way as to present the dataset to the algorithms in the most effective way possible based on its intrinsic characteristics of the algorithm. This phase was carried out using standard values of the hyperparameters that allowed us to compare the different input configurations with each other;

\item The second step concerned the optimization of the algorithms by modifying the respective hyperparameters to exploit their potential. This research on hyperparameters was carried out using exclusively the training set and the technique known as k-fold cross validation, which consists for each desired value of the hyperparameters of dividing the training set into $k$ sets and in rotation using $k-1$ to the training and the remainder for validation. Once all the possible combinations have been completed, the average performance is used as a measure of the goodness of the model and as a comparison metric for the same algorithm with different values of the hyperparameters;

\item The third, and last step, consists of quantifying the errors of each algorithm in order to be able to compare them with each other. For this phase, we used exclusively the test set and different evaluation metrics with different characteristics. For definitions and their peculiarities, we refer to the appendix \ref{App:metrics}.
\end{enumerate}

\section{Creation of the Dataset}
\label{Sec:dataset}

Each supervised learning algorithm needs a dataset to start from and, given its importance, we will focus on the creation and exploration of it. Since our goal is to predict the price of Bermudan Swaptions starting from some of their characteristics available to market participants, we need a dataset containing this information. Specifically, the prices of the Bermudan Swaptions represent the dependent variable also known as the target, while all the information that we decide to use as independent variables, is known as features. Furthermore, it is well to underline that in our case the quantity to be predicted is a single real number (the Bermudan Swaption price) and therefore the problem we face falls into the category of single output regression problems. 

For the creation of the dataset, we selected a heterogeneous set of 434 Bermudan Swaptions which we report entirely in Figure \ref{Fig:basket} of Appendix \ref{App:basket}. The Bermudan Swaptions considered have different characteristics, in fact they are both payer and receiver version with different tenor, no call period and strike. Specifically, the tenor represents the duration of the underlying swap contract, the no call period is the period until the first possible exercise date while the strike is the distance in basis points from the ATM. We report in Table \ref{Tab:feature_dataset} the assumed values of the features just mentioned. Unlike pricing these instruments in a single market scenario, we considered different market scenarios to increase the number and variability of our dataset. This operation is possible considering multiple values of the parameters of the short rate model implemented. Pricing the entire Swaption set with different speed of mean reversion and volatility values allows us to consider different variance levels of the underlying stochastic processes, thus generating different market situations. In theory, these parameters could be chosen arbitrarily, but to obtain values that were reasonable with today's market we acted differently: we calibrated the G1++ parameters for each of the Bermudan Swaptions in the basket to their natural hedges, i.e. the underlying European Swaption, not only using the market data available (Appendix \ref{App:market}), but also other two market scenarios obtained by modifying the implied Black volatility of those available. Specifically, we built high and low volatility scenarios by bumping the original volatility of $+25\%$ and $-25\%$ of its original value. In conclusion, this procedure allowed us to define reasonable ranges for the parameters of the Hull-White model:
\begin{equation}
    a \in [ -2\%,30\% ],  \qquad \sigma \in [ 0.1\%,9\% ].
\end{equation}
Within this parameter space, we identified two pathological areas that are not interesting to be explored, which are respectively the one with high speed of mean reversion and low volatility values and the opposite one, i.e. with high volatility and low speed of mean reversion. Specifically, the first combination returns an almost deterministic model as it does not have volatility while the second combination returns an explosive behaviour of the model. For these reasons, we have selected a central area in which to sample the parameters. We have selected exactly 10 parameters that we report in Appendix \ref{App:hw}. Once these values were defined, it was possible to obtain the price, through the LSMC, for each of the 434 Bermudan Swaptions in the basket for a total of 4340 prices (10 different scenarios for each Swaption). With the aim of speeding up the computation, we parallelized the LSMC code on a cluster; specifically, we used 25 CPU cores each of which are entrusted with $2 \cdot 10^{4}$ simulations for a total of $5 \cdot 10^{5}$ Monte Carlo paths for each sample.

\begin{figure}[t]
\centering
\includegraphics[width=0.45\linewidth]{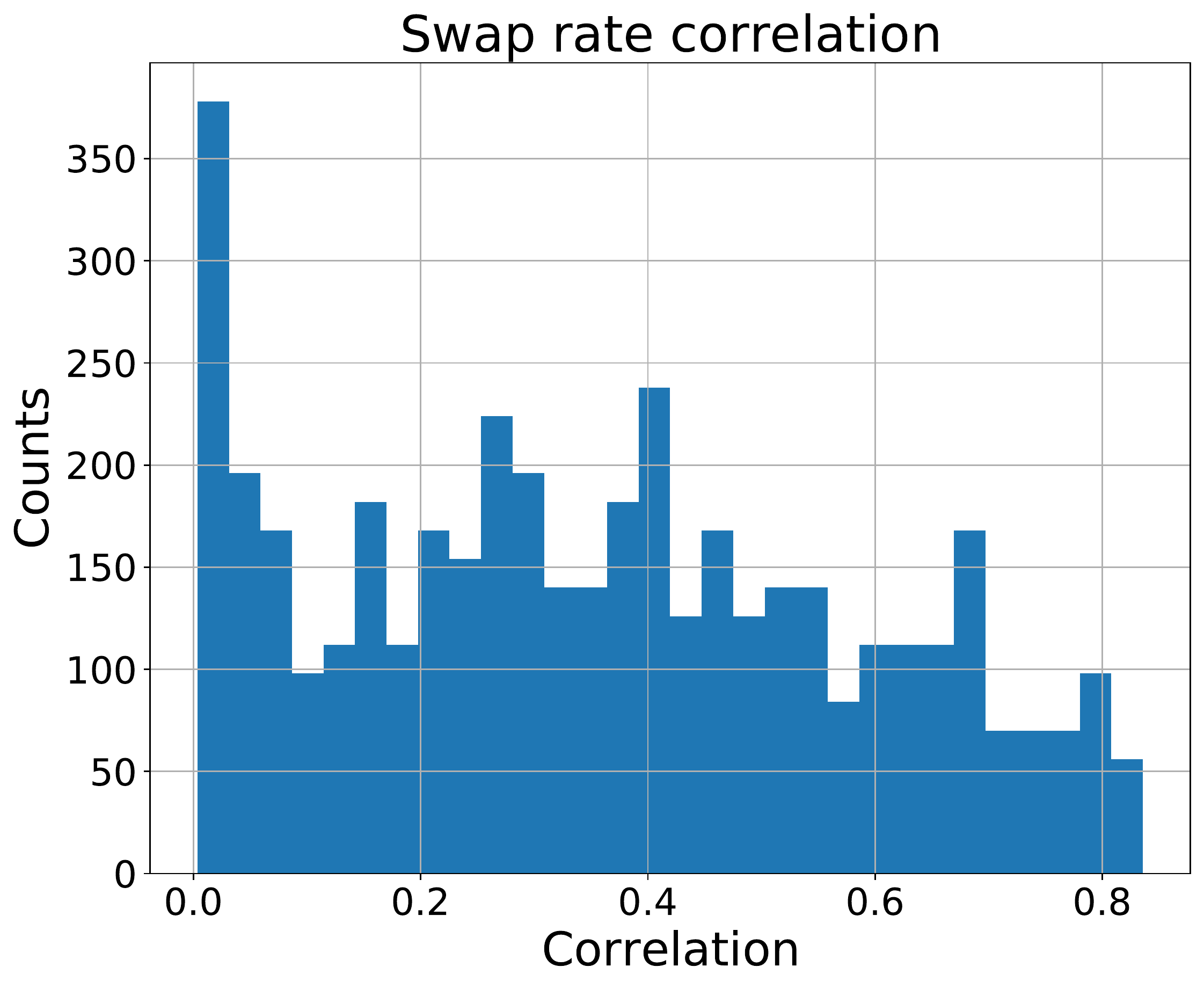}
\includegraphics[width=0.45\linewidth]{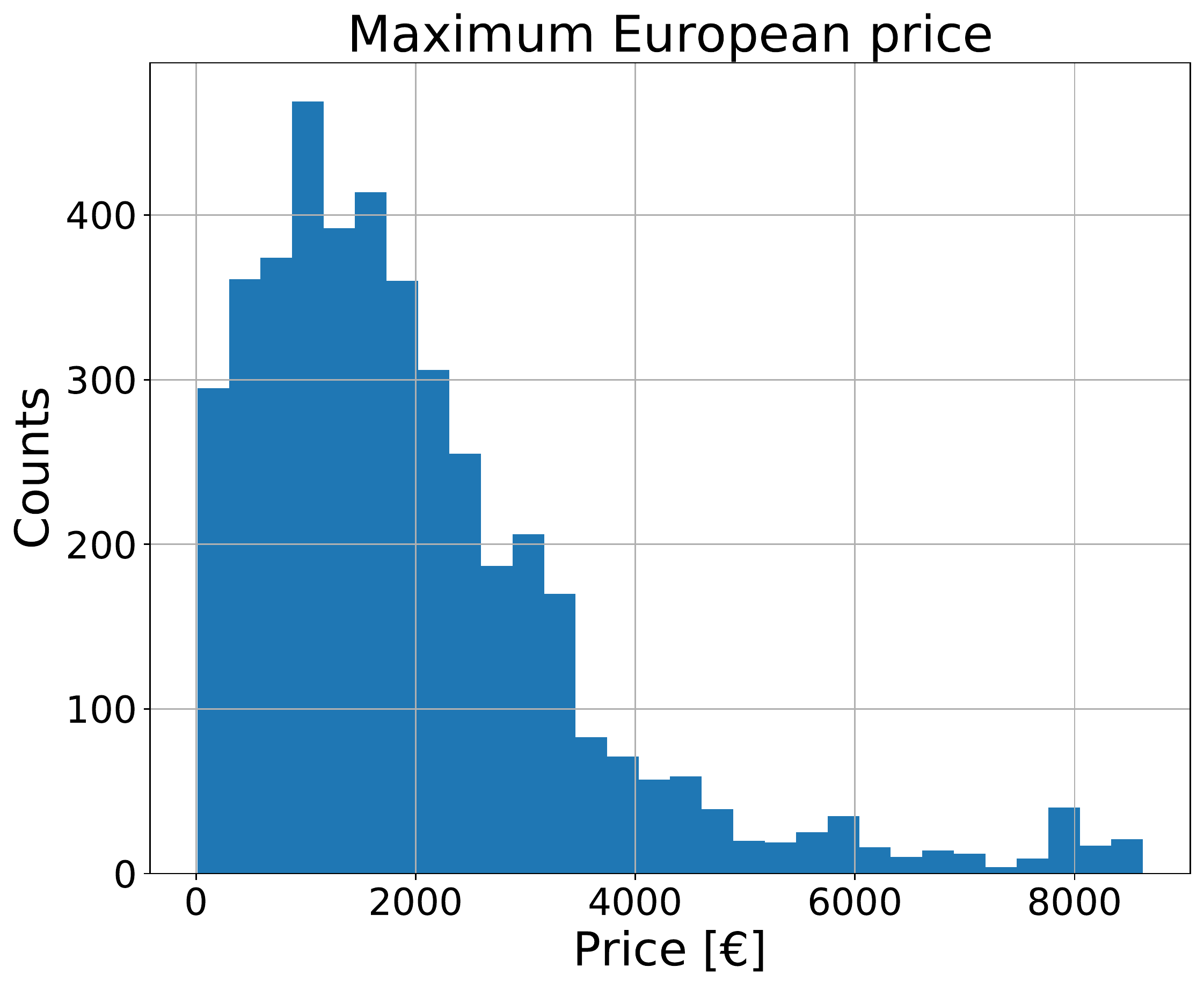}
\caption{Distribution of the correlation between the swap rates (left) and the price of the maximum European Swaption (right). It can be noted that the correlations obtained cover the space in a homogeneous way while the prices of European Swaptions, obtained with the closed formula of G1++, vary on very different scales.}
\label{Fig:corr_europ}
\end{figure}

\begin{figure}[t]
\centering
\includegraphics[width=0.45\linewidth]{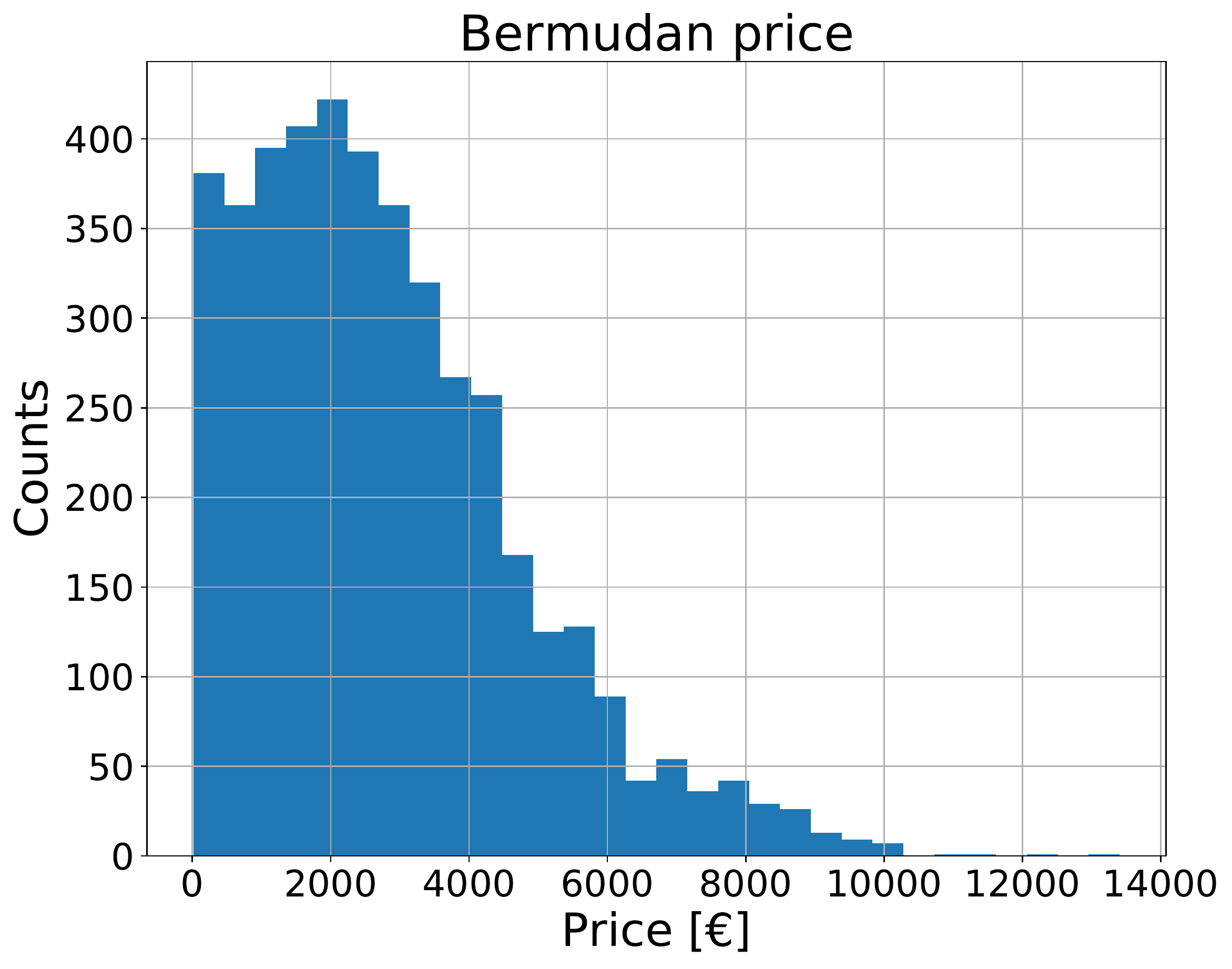}
\caption{Distribution of the target (Bermudan price). The price of the Bermudan options is obtained through the Least Square Monte Carlo algorithm with  $5 \cdot 10^{5}$ paths each. Note that the price obtained varies on very different scales.}
\label{Fig:target}
\end{figure}

Having defined the possible values of the G1++ parameters and obtained the corresponding price, i.e. the target, now we just have to identify the features. Since we want SL algorithms to be independent of the underlying model, neither the speed of mean reversion nor the volatility will be used as a feature, but we have decided to designate as independent variables the contractual information of the Bermudan Swaptions in the basket, such as the tenor, the strike, the side and the no call period. This information uniquely identifies the 434 Bermudan Swaptions that make up our basket. We have decided not to include maturity as a feature because, knowing the tenor and the no call period, it is redundant information and we have also excluded the exercising frequency as for all Swaptions it is the same (annual). To help SL algorithms to distinguish the Bermudan Swaptions in the different market scenarios, we decided to provide two additional elements that could be useful for characterising the target. First, the price of the underlying maximum European Swaption, computed with the closed-form of G1++, since we know it to be the lower bound of the Bermudan Swaption price. As second we have computed a correlation between the swap rates of the underlying European Swaptions as we know that the value of the speed of mean reversion is precisely linked to it. Specifically, we have calculated the correlation between the swap rates of the European Swaption with the longest tenor and that with the shortest tenor. For example, if we consider a Bermudan Swaption with a 10-years no-call period and a 5-years tenor, we have evaluated the correlation between the swap rates of the 11x4 European Swaption and the 14x1 European Swaption. We report in Figure \ref{Fig:corr_europ} the correlation obtained between the swap rates and the price of the maximum European Swaption while in Figure \ref{Fig:target} we report the distributions of the target (Bermudan price). 
In conclusion, to summarize, we report in Table \ref{Tab:feature_dataset} all the features (independent variables) and the target (dependent variable) that make up the dataset.
\begin{table}[t]
\centering
\begin{tabular}{cc}
\toprule
    \textbf{Features} & \makecell{tenor $\in \{2Y,5Y,10Y,15Y,20Y\}$ \\ strike $\in \{ -100,-75,-60,-50,-40,-30,-25,-20,-15,-10,-7,-5,-2,$\\$0,20,25,30,50,100,200,300,400 \} $ \\ side $\in \{\text{payer}, \text{receiver}\}$\\ no call period $ \in \{1Y,2Y,3Y,4Y,5Y,7Y,10Y,15Y,20Y\}$\\ correlation (swap rates) \\ maximum European price}  \\
\midrule
    \textbf{Target} & bermudan price \\
\bottomrule
    \end{tabular}
    \caption{Features (independent variable) and target (dependent variable) of the dataset. The total number of features selected for this issue is 6, of which 4 are contractual information related to the Bermudan Swaption while the other two are additional features. The target is unique and is represented by the price of the Bermudan Swaption obtained through the LSMC.}
    \label{Tab:feature_dataset}
\end{table}
As stated previously, in the development of SL algorithms, it is fundamental how features are presented. The side feature in our dataset is the encoding of a categorical variable to distinguish the payer version from the receiver. The most common way to represent categorical variables is one-hot-encoding; since any of the possibilities excludes the other, we have decided to create a single feature that takes value 1 when it is payer and 0 otherwise.

At this point, before applying the different SL algorithms it is necessary to separate our dataset into the training set, used to build our model, and the test set used to assess how well the model works. We decided to use 80\% (3472 samples) of the dataset for training and the remaining 20\% (868 samples) for testing. Since the data were collected sequentially before dividing the dataset it is necessary to shuffle it to make sure the test set contains data of all type. Moreover, a purely random sampling method is generally fine if the dataset is large enough, but if it is not, there is the risk of introducing a significant sampling bias. For this reason, we have performed what is referred to as stratified sampling: since we know that price of the maximum European Swaption is an important attribute to predict the price of Bermudan Swaptions, we have divided the price range of the maximum European Swaptions into subgroups and the right number of instances are sampled from each of them to guarantee that the test set is representative of the overall population. The test set thus generated has been put aside and will be used only for the final evaluation of each model. The construction of the various models and the choice of hyperparameters was based exclusively on the training set.

\section{Numerical Results}
\label{Sec:results}

\begin{figure}[t]
    \centering
    \subfigure{\includegraphics[width=0.8\linewidth]{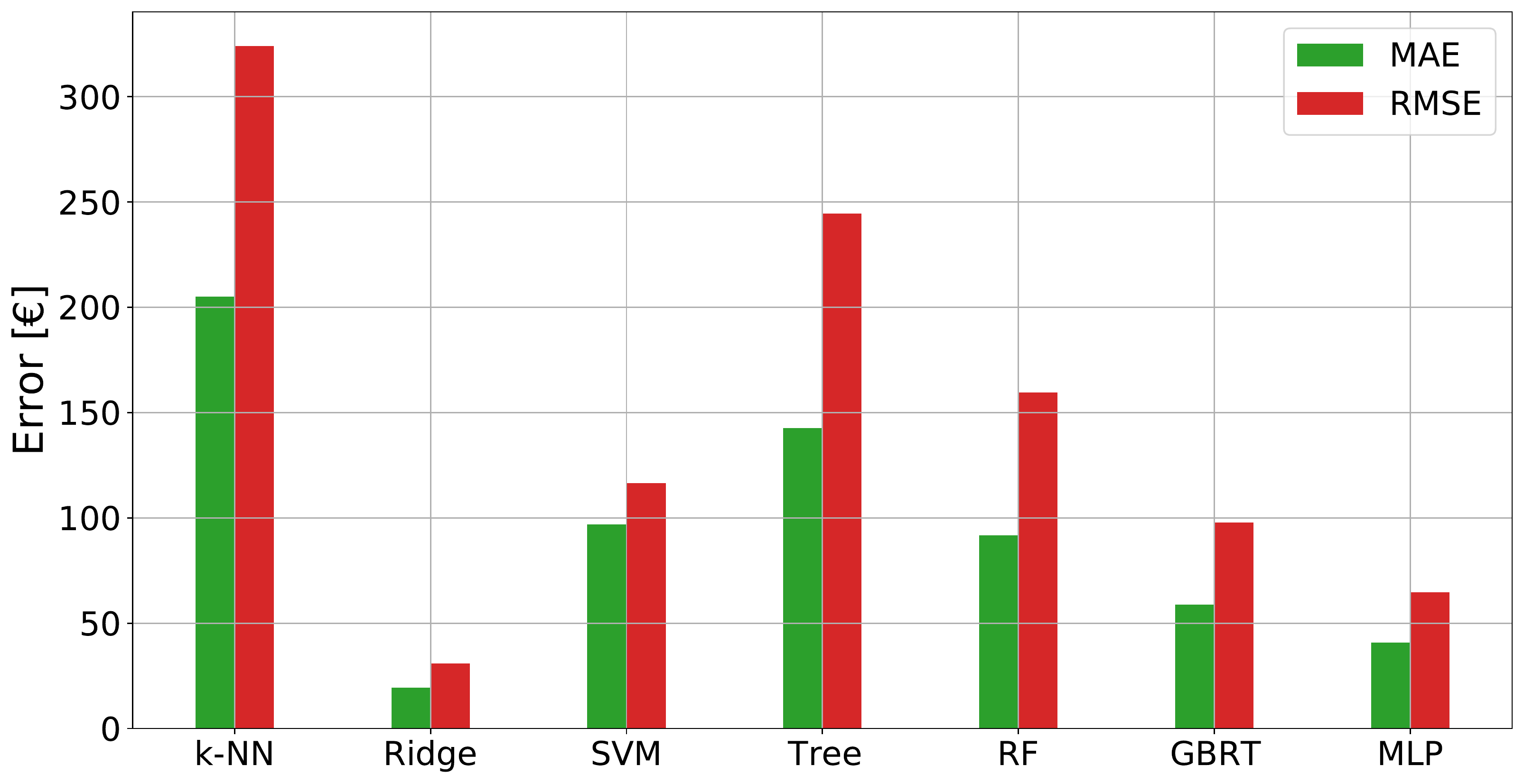}}
    \subfigure{\includegraphics[width=0.8\linewidth]{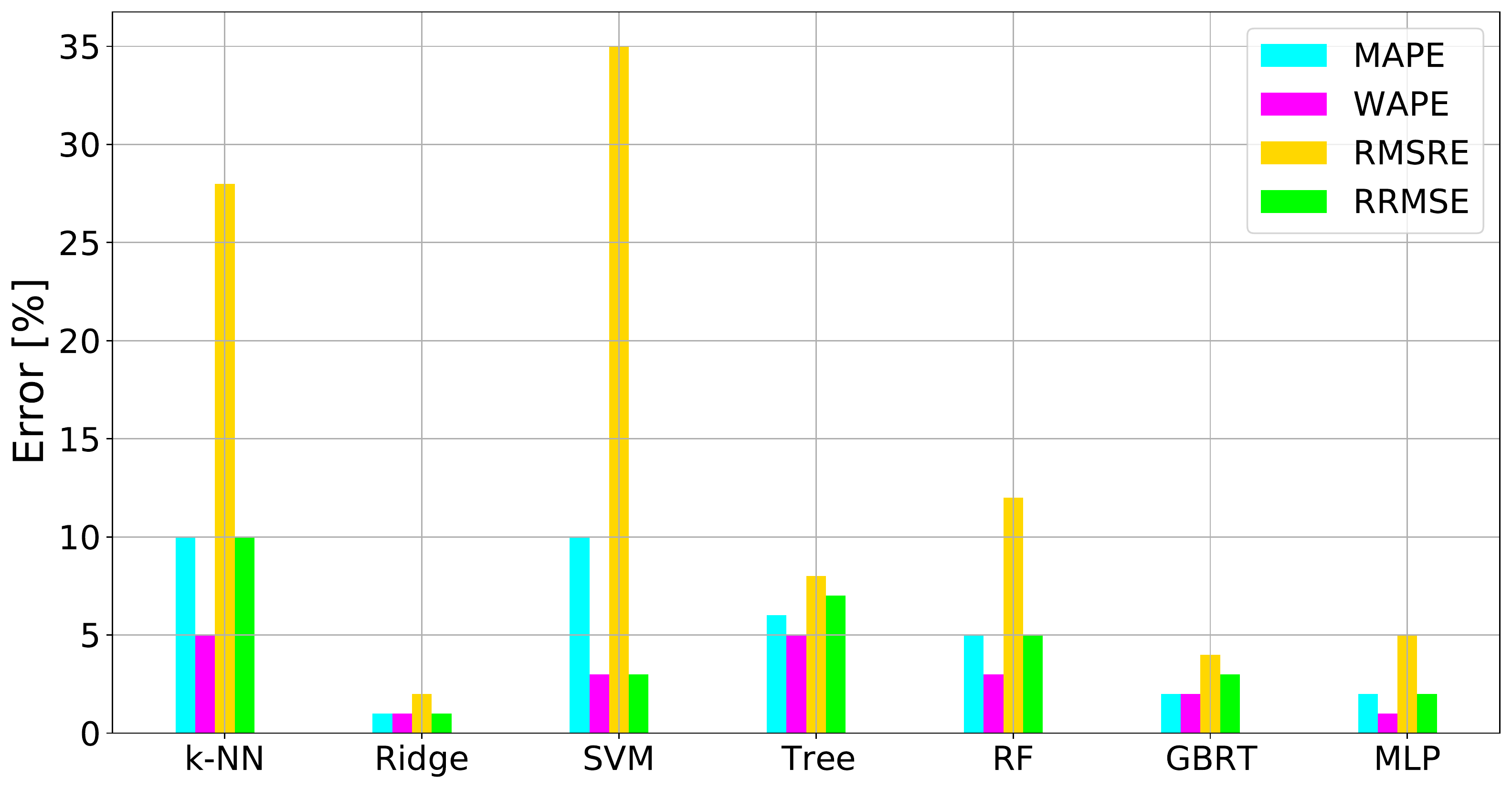}}
    \captionof{figure}{Values of the different evaluation metric on test set for each algorithm. The graph above shows the comparison for the absolute metrics, that is, those that report the error in the unit of interest (euro). The graph below shows the comparison for relative metrics in which the relative error is expressed in percentage terms.}
    \label{Fig:comparison_metric}
\end{figure}

\begin{figure}[t]
    \centering
    \subfigure{\includegraphics[width=0.6\linewidth]{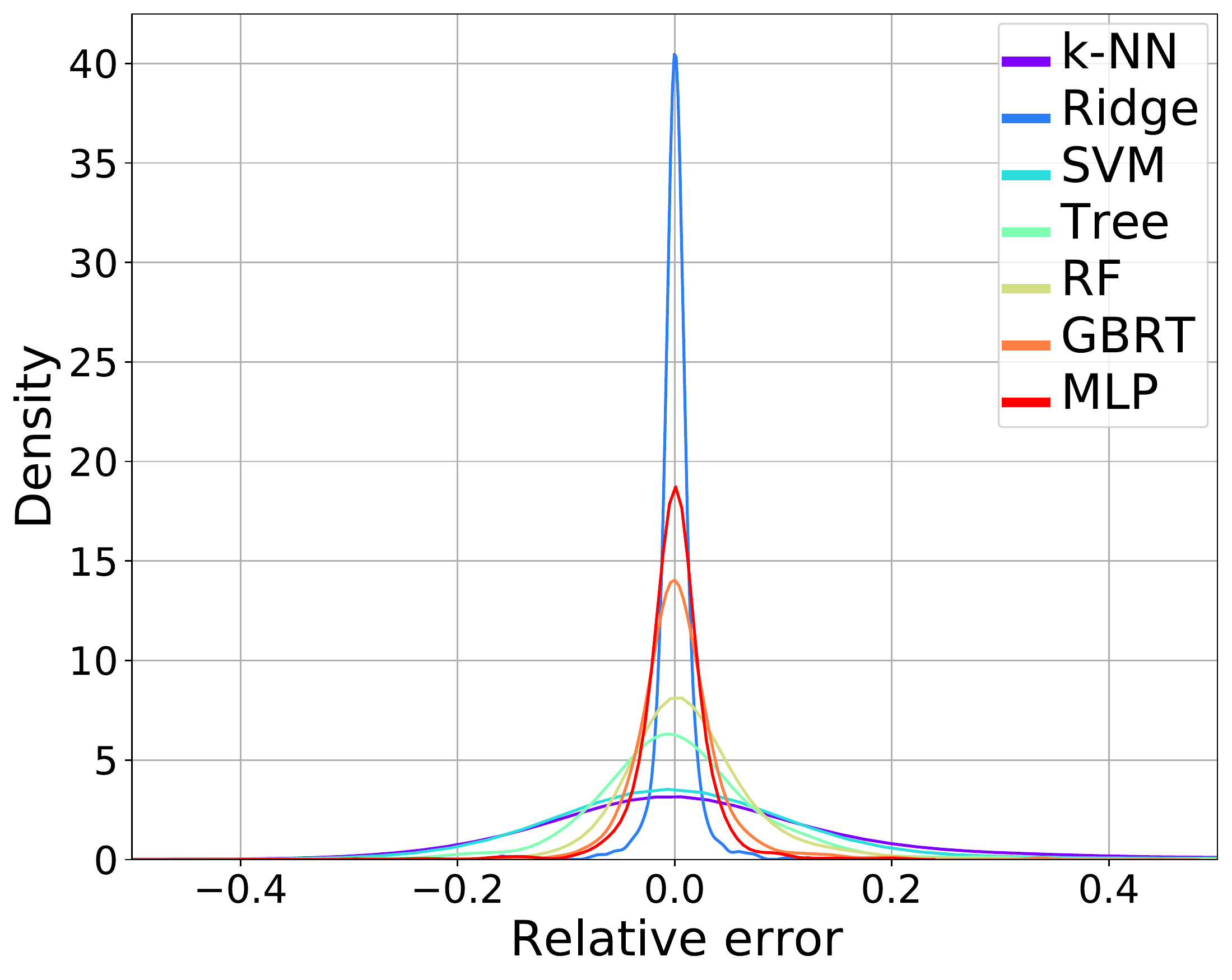}}
    \captionof{figure}{Relative error distributions for each of the algorithms. To make them comparable, all the distributions were superimposed and the interval was reduced and for this reason some of the distribution queues are not visible. The distributions were obtained with a kernel density estimation.}
    \label{Fig:comparison_distribution}
\end{figure}

This section is devoted to the comparison of the predictive performance of all the algorithms analyzed. For simplicity, we will not report here the data preparation and optimization phase of the individual algorithms, but we report in Table \ref{Tab:optimized_hyperparameters}  all of them with their respective pre-processing phase and the optimized hyperparameters on the training set. All the algorithms were implemented through Python open-source libraries like \href{https://scikit-learn.org/stable/}{scikit-learn}, \href{https://keras.io}{Keras} and \href{https://www.tensorflow.org}{Tensorflow} on a MacBook Pro (MacOS version 10.15.7) with an Intel Quad-Core i5 2.3 GHz processor with a memory of 2133 MHz and 8GB of RAM.
\begin{table}[t]
\centering
    \begin{tabularx}{\textwidth}{Xccc}
    \toprule
    \textbf{Algorithm} & \textbf{Pre-processing} & \textbf{Hyperparameters} \\ 
    \Xhline{2\arrayrulewidth}
    \textbf{k-NN} & Feature standardization & \{\texttt{k=4}\}\\
    \hline
    \textbf{Ridge}  &  \makecell{Feature standardization; \\ polynomial features }  & \{\texttt{alpha}=0.01, \texttt{degree}=6\}\\
    \hline
    \textbf{SVM}  & Feature and target standardization & \makecell{\{\texttt{kernel}=rbf, \texttt{C}=100,\\ \texttt{gamma}=0.1\}}\\
    \hline
    \textbf{Tree} & / & \{\texttt{max\_depth}=11, \texttt{min\_samples\_leaf}=0.01\}\\
    \hline
    \textbf{RF} & / & \makecell{\{\texttt{max\_depth}= 17, \texttt{max\_features} = log2,\\ \texttt{n\_estimators} = 500\}}\\
    \hline
    \textbf{GBRT} & / &  \makecell{\{\texttt{learning\_rate}=0.1, \texttt{n\_estimators}=1000, \\ \texttt{max\_depth}=5, \texttt{max\_features}=log2\}} \\
    \hline
    \textbf{MLP} & Feature and target standardization & \makecell{\{\texttt{inizialization}=He normal, \\ \texttt{activation\_function}=ReLU, \\ \texttt{optimizer} = Nadam\\ \texttt{batch\_size}=32, \texttt{learning\_rate}=0.01, \\ \texttt{n\_hidden} =4, \texttt{n\_neurons}=100 \}}\\
    \bottomrule
  \end{tabularx}
  \caption{All models with their pre-processing phase and optimized hyperparameters. We have standardized features by removing the mean and scaling to unit variance; centering and scaling happen independently on each feature/target by computing the relevant statistics on the samples in the training set.}
\label{Tab:optimized_hyperparameters}
\end{table}
An easy way to compare models and their predictive capabilities is to observe their performance on the test set. For this purpose, we report in Figure \ref{Fig:comparison_metric} the comparison between all the values of the evaluation metrics (Appendix \ref{App:metrics}), both absolute and relative, grouped by the algorithm. For completeness, we also report in Table \ref{Tab:all_statistics} a comparison between all the indices of the relative error distribution for each of the algorithms and in Figure \ref{Fig:comparison_distribution} the respective error distributions. Furthermore, we have also reported in Table \ref{Tab:computation_time} the comparison between the training and pricing times for all the supervised algorithms of 434 Bermudan Swaptions. To compare these results with the standard method, we priced the same set with the LSMC considering $5 \cdot 10^{4}$ path for each Bermudan Swaption obtaining a pricing time equal to 1086.6 seconds.
Given these results, the first observation in Figure \ref{Fig:comparison_metric} is purely statistical; as expected the RMSE values are always greater than the MAE ones for all the algorithms. Furthermore, the values of WAPE and RRMSE, introduced to reduce and limit some negative aspects of the MAPE and the RMSRE respectively, are in fact lower or equal to the latter. Furthermore, it can be observed that the model that can be considered the worst for this type of problem is undoubtedly the k-NN as it has the highest generalization error in almost all the metrics considered. We believe that this is due to the too-simple nature of the algorithm and above all to the lack of flexibility of its hyperparameters, which limit the reachable complexity. Among all the tree-based models, we can observe that the RF and GBRT perform better than the simple decision tree as we reasonably expect for ensemble methods. The best of this kind of model and the most promising is the GBRT that have the lowest generalization error of all and for all the metric considered. The great strength of this type of algorithm, which makes them very versatile, is the fact that they require practically no preprocessing of the data. For this reason, we consider the GBRT promising and usable even with a larger dataset above all as the first-entry algorithm. Instead, SVM has slightly worse performance than the GBRT for all the metrics considered; also note that it has the highest RMSRE value among all the analyzed models. In conclusion, let us consider the two best algorithms obtained. The best performance of all belongs to the Ridge regression. Moreover, note that it has the lowest generalization error whatever the metric considered. A slightly worse result than this, but still very good, is obtained by MLP. However, we believe that with even more research on hyperparameters and especially with a greater amount of training data, ANN could improve their performance. The only downside to the MLP compared to the Ridge is the long time it takes to train the model (Table \ref{Tab:computation_time}). 
\begin{table}[t]
\centering    
    \begin{tabularx}{\textwidth}{Xcccccccc} 
\toprule
      & \textbf{k-NN} & \textbf{Ridge} & \textbf{ SVM} & \textbf{Tree} & \textbf{RF} & \textbf{GBRT} & \textbf{MLP}\\ 
\midrule
    \textbf{mean} & 0.0524 & 0.0006 & 0.0475 & 0.0329 & 0.0243 & 0.0036 & 0.0002\\
  \textbf{std} & 0.2771 & 0.0182 & 0.3422 & 0.8381 & 0.1153 & 0.0444 & 0.0041\\
    \textbf{skew} &  6.6101 & -0.2800 & 9.6765 & 0.7100 & 7.2053 & 0.3922 & 1.2675\\
    \textbf{kurtosis} & 66.9179 & 24.2352 & 126.0146 & 5.6378 & 71.2143 & 17.3850 & 58.2962\\
    \textbf{min} & -0.4514 & -0.1608 & -0.8059 & -0.3820 & -0.2713 & -0.4172 & -0.4506\\
    \textbf{25\%} & -0.0517 & -0.0049 & -0.0344 & -0.0402 & -0.0154 & -0.0157 & -0.0128\\
    \textbf{50\%} &  -0.0053 & -0.0004 & -0.0029 & -0.0014 & 0.0049 & 0.0001 & -0.0013\\
   \textbf{75\%} &  0.0747 & 0.0061 & 0.0353 & 0.0412 & 0.0363 & 0.0191 & 0.0119\\
 \textbf{max} &  3.9007 & 0.1549 & 5.2834 & 0.5939 & 1.5604 & 0.3683 & 0.5659\\
    \bottomrule
  \end{tabularx}
  \caption{Comparison for all relevant statistics of relative error for each algorithm. In addition to means and standard deviation, the third (skewness) and fourth (kurtosis) moments are also reported. The quantiles, i.e. the value below which a certain percentage of the errors was found, are also reported between the minimum value and the maximum value obtained.}
\label{Tab:all_statistics}
\end{table}
All these deductions are also supported by the information reported in Table \ref{Tab:all_statistics} and Figure \ref{Fig:comparison_distribution}. In fact, it can be noticed the three algorithms that have been identified as the best, have the average values closest to zero with the lowest standard deviation. Furthermore, it can also be seen from the values of skewness, kurtosis and quantiles that these models are characterized by the most symmetrical distributions without outliers. All the others, on the other hand, are characterized by higher standard deviations and in some cases larger tails of the distributions.
\begin{table}[t]
\centering
    \begin{tabular}{ccc}
    \toprule
    \textbf{SL Algorithm} & \textbf{Training Time} &  \textbf{Pricing Time} \\ 
    \midrule
    \textbf{k-NN} & $12.9 \cdot 10^{-3}$ s & $9.4 \cdot 10^{-3}$ s\\
    \textbf{Ridge} &  $316 \cdot 10^{-3}$ s & $7.7 \cdot 10^{-3}$ s\\
    \textbf{SVM} & $736 \cdot 10^{-3}$ s & $161 \cdot 10^{-3}$ s\\
    \textbf{Tree} & $15.9 \cdot 10^{-3}$ s & $4 \cdot 10^{-3}$ s\\
    \textbf{RF} &  2.1 s & $71.3 \cdot 10^{-3}$ s\\
    \textbf{GBRT} &  1.6 s & $10.9 \cdot 10^{-3}$ s\\
    \textbf{MLP} &  28 s & $98.7 \cdot 10^{-3}$ s\\
    \bottomrule
    \end{tabular}
    \caption{Comparison between training and pricing times of all supervised algorithms. The pricing times take into consideration 434 Bermudan Swaptions while the training was carried out with 3472 Swaptions.}
    \label{Tab:computation_time}
\end{table}
In general, from Figure \ref{Fig:comparison_metric} it can be seen, apart from a few exceptions, that the result of the comparison between two models does not change if we observe different metrics. In other words, if one model is better than another by considering the error reported by one metric, it will remain better even if they are compared using a different metric. Consequently, if the goal is the pure comparison between models, we can say that the use of a particular metric with respect to another is useless. The use of a particular metric becomes decisive if we consider the purpose of the work and what represents the generalization error. Since in our case the goal was to predict prices over an extended range with different scales, we believe a \virgolette{relative} metric has more meanings than an \virgolette{absolute} one. Specifically, among the \virgolette{relative} metrics (bottom of Figure \ref{Fig:comparison_metric}), we prefer the use of RRMSE (\virgolette{lime} color) 
for its intrinsic characteristics. In conclusion, we can say that the average price error of the Ridge equal to 1\% is an excellent result in comparison to 2\% average of the standard deviation found from market data. 

\begin{figure}[t]
    \centering
    \subfigure{\includegraphics[width=0.95\linewidth]{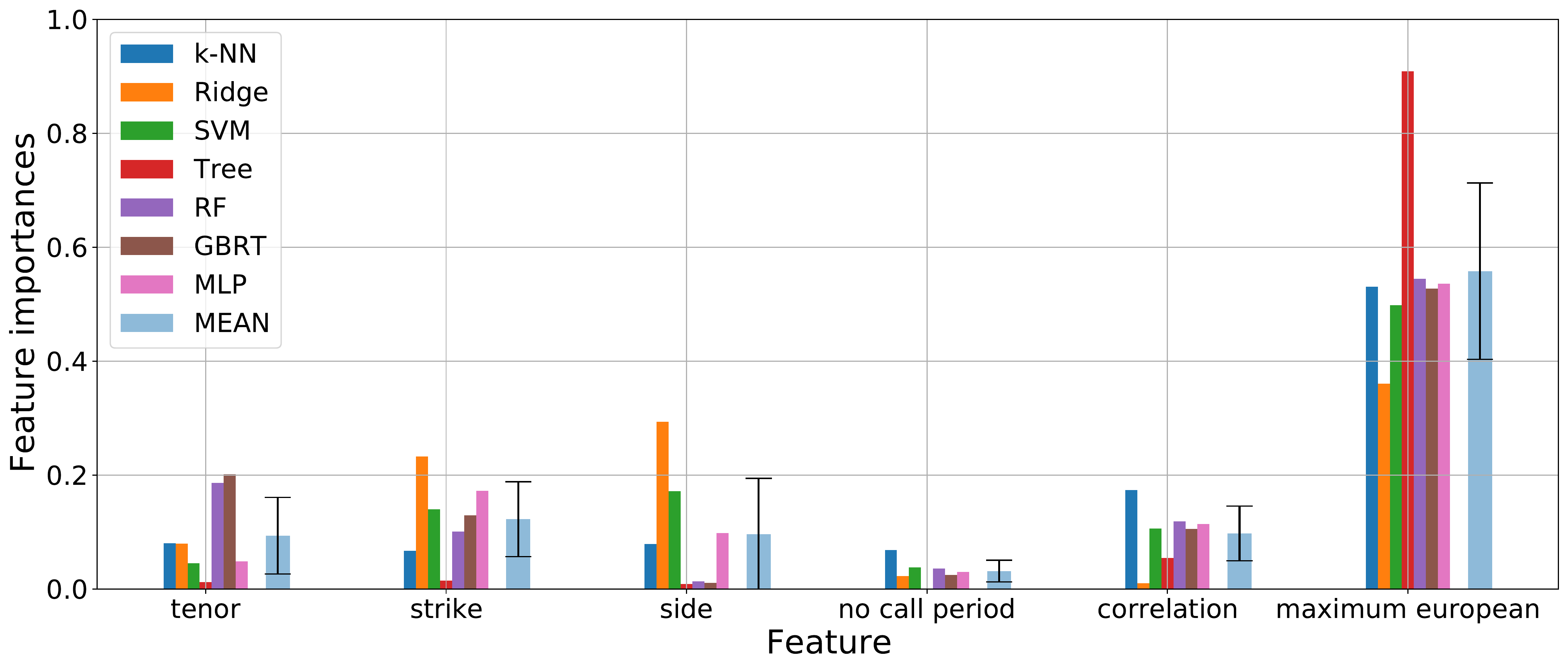}}
    \captionof{figure}{Feature importance for each of the algorithms grouped by feature. The values assigned by the algorithms to each of the features have been normalized so that sum up to 1. Note that all tree-based models have an endogenous method while for all the other models we used an indirect method, known as permutation importance. The last bar on the right for each feature represents the mean value and standard deviation of the feature importance assigned by each of the individual algorithms.}
    \label{Fig:feature_importances}
\end{figure}

Typically, in SL algorithms it is customary to ask what is the relative weight of the independent variables in target prediction and this analysis is commonly known as feature importance. In other words, it gives a qualitative measure of the impact that each explanatory variable has in predicting the target. In Figure \ref{Fig:feature_importances} for each of the features, the importance assigned by each of the algorithms is reported together with their average value and the standard deviation.
Specifically, all methods based on decision trees have an endogenous method which is based on the reduction of the value of the metric used in the construction of the trees. For all the other models, however, we used an indirect method, known as permutation importance, which consists of evaluating the deterioration in performance when the values of a feature are randomly mixed and then using it as an indirect measure of the importance of a variable.
  
From the Figure \ref{Fig:feature_importances} we can see a significant aspect: although with different weights, all the models developed indicate the price of the maximum underlying European Swaption as the most explanatory variable. This outcome is reassuring as all models are able to recognize that the price of the Bermudan Swaption is closely linked to the price of the maximum European Swaption which constitutes its lower bound.  Furthermore, except for the no call period which is practically unused by all algorithms, the other features have comparable average values, with the only difference that the correlation between the swap rates has the lowest standard deviation, a sign that the returned weights by the individual algorithms are very similar to each other.

\section{Conclusion and Future Works}

In this final part, we want to retrace the fundamental stages of this work focusing on the main results and developments presented and pointing out some possible prospects. This project was devoted to the use of SL algorithms to solve optimal stopping time problem. In particular, we focussed on the pricing of Bermudan Swaptions, where we thought such techniques could be helpful to overcome the computational bottleneck of dynamic programming algorithms and to understand the most important driving factors behind the market price. Moreover, we want to emphasize the fact that this work is fully extendable to any other American-type financial product.

First, we implemented a Hull-White One Factor model (G1++) that was used to generate scenarios for the evolution of the interest rate curve. Then we developed the Least Squares Monte Carlo engine, which is one of the most widely used numerical algorithms to price Bermudan Swaptions. Thanks to these tools we have created a reasoned dataset containing prices of 434 different Bermudan Swaptions in 10 different market scenarios, for a total of 4340 data points. To find out which SL methodology best suited our problem, we have focussed on a subset of quite different algorithms and applied them using the optimized Python libraries for ML. To assess the  predicting capabilities of each algorithm, our approach was based on three stages. The first consisted of pre-processing data and performing feature engineering to present the dataset to our algorithms in the most effective way possible. The second concerned the fine-tuning of the hyperparameters to get optimized predictors. The third and last is focused on quantifying the fitting performances of our models. We concluded our work by presenting a comparative analysis of all implemented SL algorithms. From this last comparison, we could assert that the use of these techniques for the pricing of Bermudan Swaptions is very promising as the prices predicted by the algorithms were very close to the real ones. Especially they overcome the computational bottleneck of numerical simulations, in fact it can be seen from Table \ref{Tab:computation_time} that each of the supervised algorithms considered has  a pricing time of at least 6 orders of magnitude lower than a traditional dynamic programming algorithm. Two algorithms with excellent performances emerged from our study: ANN and Ridge Regression. Among these, the Ridge has the advantage of having one of the fastest training phases, while the ANN is the slowest and requires a more accurate tuning of its hyperparameters. A further promising algorithm was the GBRT, especially for minimal data preparation required and the intrinsic method of evaluating the importance of features. Moreover, thanks to feature importance techniques we have confirmed that the most determining factor for the price of a Bermudan Swaption is the value of the maximum underlying European Swaption, which constitutes its lower bound. 

The results obtained in this work open up several possibilities for prospects. First of all, the generation of the dataset could be extended and refined in two different ways: either in the direction of numerosity, thus increasing the number of options in the basket or the number of market scenarios to study the change in the performance of the algorithms, or by building a dense dataset but in market scenarios with a very similar variance, such that the correlation between swap rates should have greater importance. In our framework, correlations between co-terminal rates could be explicitly calculated and were input of the SL model. Unfortunately, it is very difficult to infer this kind of information from the market, and our approach could also prove helpful the other way round, i.e. to estimate correlations from recent historical Bermudan prices. Correlation estimates would then be used straightforwardly as the input of the algorithms implemented in this project.

\newpage
%--- Appendices
\begin{appendices}

\section{SL algorithms}
\label{App:SL}

We present a list of the SL algorithms chosen in this work and their main characteristics and differences.
\begin{itemize}
\item \textbf{k-Nearest Neighbour (k-NN)}. This algorithm is arguably the simplest, however being a non-parametric algorithm, i.e. it does not make assumptions regarding the dataset, it is widely used. The principle behind nearest neighbour methods is to find a predefined number $k$  of training samples closest in distance to the new point and predict the label from these. As there is a data storing phase, it is well suited for a small dataset (both in the number of features and in the number of samples) and it is known not to work well on sparse data\footnote{Few data in large hyper-volume, i.e. most are zero.} and and features must have the same scale since absolute differences must weight the same. The label assigned to a query point is computed based on the mean of the labels of its nearest neighbours. The model mainly presents three important hyperparameters: the number of neighbours k, the metric used to evaluate the distance and the weights assigned to the neighbours to define their importance.

\item \textbf{Linear Models}. Linear models are a class of models widely used in practice because they are very fast to train and predict; they make a prediction using a linear function of the input features, i.e. the target value is expected to be a linear weighted combination of the features. Notice that the linearity is a strong assumption and it is not always respected, but this gives them an easy interpretation. Training a model like that means setting its parameters so that the model best fits the training set. In general, linear models are very powerful with large datasets, especially if the number of features is huge (high-dimensional problem). There are many different linear models and the difference between them lies in how the parameters are learned and how the model complexity can be controlled. We have considered the \textbf{Linear Regression} and two of its regularised versions: \textbf{Ridge} and \textbf{Lasso Regression} where the regularization term is respectively the $L^{2}$ and the $L^{1}$ norm of the weight vector.

\item \textbf{Support Vector Machine (SVM)}. Conceptually the SVM, using some significant data points (support vector), try to define a \virgolette{corridor} (or a hyper-volume in higher dimensions) within which the greatest number of data points fall. In general, SVMs are effective in the higher dimension, but they do not perform well when we have a large dataset due to the higher training time. They have a hyperparameter that performs the same task as alpha of the linear models and therefore it limits the importance of each support vector. SVMs are efficient also for non-linear problems thank to a mathematical technique called kernel trick; depending on the kernel used, additional hyperparameters are needed, but we must take that into account that one of the biggest drawbacks of these algorithms is the high sensitivity to hyperparameters.

\item \textbf{Tree-based algorithms}. As the name suggests, these algorithms are based on simple decision trees. Like SVMs, decision trees are versatile and very powerful and like K-NN, they are non-parametric algorithms. The goal of these algorithms is to create a model that predicts the value of a target variable by learning simple decision rules inferred from the data features. To build a tree, the algorithm searches all over the possible tests (a subdivision of the training set) and find the one that is most informative about the target variable. This recursive process yields a binary tree\footnote{We have considered only binary trees.}, with each node containing a test and it is repeated until each region in the partition only contains a single target value. A prediction on a new data point is made by checking which partition of feature space the point lies in and the output is the mean target of the training point in this leaf. One of the many qualities of decision trees is that they require very little data preparation, moreover, they are very fast to predict and they are defined as \virgolette{white model} because they are easily interpretable. Typically, building a tree and continuing until all leaves are pure leads to models that are very complex and highly overfit to the training data and therefore they provide poor generalization performance. The most common way to prevent overfitting is called pre-pruning and it consists of stopping the creation of the tree early. Possible criteria for pre-pruning include limiting the maximum depth of the tree, limiting the maximum number of leaves and others making the decision trees highly dependent on the numerous hyperparameters. Moreover, they have two main problems: the first is the inability to extrapolate or make predictions outside the training range, while the second is that they are unstable due to small variations in the training set. This last problem is solved with the decision tree ensembles: \textbf{Random Forest (RF)} and \textbf{Gradient Boosted Regression Tree (GBRT)}. 

Ensembles are methods that combine multiple SL models to create a more powerful one. They are based on the idea that the aggregation of the predictions of a group of models will often give better results than with the best individual predictor. One way to obtain a group of predictors is to use the same algorithm for every model and train them on different random subsets of the training set; when sampling is performed with replacement, this method is called bagging, otherwise, it is called pasting. Generally, the net result is that the ensemble has a similar bias but a lower variance than a single predictor trained on the original training set. Below we briefly describe the two ensemble methods considered.
\begin{itemize}
\item A \textbf{RF} is an ensemble method of decision trees, generally trained via bagging method \cite{Bre01}. The idea behind random forests is that each decision tree will likely overfit on a specific part of the data, but if we build many trees that overfit in different ways, we can reduce the amount of overfitting by averaging their results. RF get their name from injecting randomness into the tree building in two ways: by bagging method and by selecting a random subset of the features in each split test. In summary, the bootstrap sampling leads to each decision tree in the RF being built on a slightly different dataset and due to the selection of features in each node, each split in each tree operates on a different subset of features. Together, these two mechanisms ensure that all the trees in the RF are different. Essentially, RF shares all pros of the decision tree, while making up for some of their deficiencies; it also has practically all their hyperparameters with the addition of a new one that regulates the number of trees to consider whose greater values are always better, because averaging more trees will yield a more robust ensemble by reducing overfitting.

\item \textbf{GBRT} \cite{Fri02} is part of the more general boosting method in which predictors are trained sequentially, each trying to correct its predecessor. By default there is no randomization in gradient boosted decision trees instead, strong pre-pruning is used; it often uses very shallow trees which makes the model smaller in terms of memory and makes predictions faster. Each tree can only provide good predictions on part of data, and so more and more trees are added to iteratively improve performance. This method shares the same hyperparameters as RF with the addition of the learning rate but, in contrast to RF, increasing the number of predictors leads to a more complex model. The learning rate and the number of estimators are highly interconnected, as a lower rate means more trees are needed to build a model of similar complexity and therefore there is a trade-off between them. Similar to other tree-based models, the GBRT works well without scaling and often does not works well on high-dimensional sparse data. Their main drawback is that they require careful tuning of hyperparameters and may take a long time to train.
\end{itemize}

\item \textbf{Artificial Neural Networks (ANN)} or \textbf{Multi-Layer Perceptron (MLP)}. They can be understood as a large set of simpler units, called neurons, connected in some way and organized in layers. An ANN is composed of one input layer, one or more hidden layers and one final output layer. In order to understand the entire functioning of the network, it is necessary to consider a single neuron: the inputs and the output are numbers and each input connection is associated with a weight. The artificial neuron computes a weighted sum of its inputs and then applies a non-linear transformation, called activation function. In some way, ANNs can be viewed as generalizations of linear models that perform multiple stages of processing to come to a decision. The key point of ANN is the algorithm used to train them; it is called back-propagation algorithm and in simple terms, it is a gradient descent using an efficient technique for computing the gradients automatically. In conclusion, ANNs are typically black box models defined by a set of weights; they take some variables as input and modify the values of the weights so that they return the desired target. Given enough computation time, data and careful tuning of the hyperparameters, ANN are the most powerful and scalable ML models. The real difficulty in implementing a suitable model is contained in the enormous amount of hyperparameters that regulate the complexity of the network. Both the number of hidden layers and the number of neurons in each layer can affect the performance of an ANN, but there is a large variety of hyperparameters that need to be optimized for acceptable results. In general, choosing the exact network architecture for an ANN remains an art that requires extensive numerical experimentation and intuition, and is often problem-specific. 
\end{itemize}

\newpage
\section{Error Metrics}
\label{App:metrics}

We present a list of the metrics implemented and their main characteristics and differences.
\begin{itemize}
    \item \textbf{MAE}. It measures the average magnitude of the errors in a set of predictions, without considering their direction. It’s the average over sample ($n$) of the absolute differences between target ($y_{i}$) and prediction ($\hat{y}_{i}$) where all individual differences have equal weight. In formula:
        \begin{equation}
        MAE \coloneqq \frac{1}{n} \sum_{i=1}^{n}\left|y_{i}-\hat{y}_{i}\right|
        \end{equation}
    \item \textbf{MAPE}. Instead of using actual value, MAPE uses the relative error to present the result. It is defined as
        \begin{equation}
            MAPE \coloneqq \frac{1}{n} \sum_{i=1}^{n}\left|\frac{y_{i}-\hat{y}_{i}}{y_{i}}\right|
        \end{equation}
        MAPE is also sometimes reported as a percentage, which is the above equation multiplied by 100.
    \item \textbf{WAPE}. It is relative to what it would have been if a simple predictor had been used. More specifically, this simple predictor is just the average of the real values. Thus, it is defined dividing the sum of absolute differences and normalizes it by dividing by the total absolute error of the simple predictor. In formula:
        \begin{equation}
            WAPE \coloneqq \frac{\sum_{i=1}^{n}\left|y_{i}-\hat{y}_{i}\right|}{\sum_{i=1}^{n}\left|y_{i}\right|}
        \end{equation}
        WAPE is also sometimes reported as a percentage, which is the above equation multiplied by 100.
    \item \textbf{RMSE}. It represents the square root of the second sample moment of the differences between predicted values and real values. In formula:
        \begin{equation}
            RMSE \coloneqq \sqrt{\frac{1}{n} \sum_{i=1}^{n} \left(y_{i}-\hat{y}_{i}\right)^{2}}
        \end{equation}
    \item \textbf{RMRSE}. It is defined as
        \begin{equation}
            RMSRE \coloneqq \sqrt{\frac{1}{n} \sum_{i=1}^{n} \left(\frac{y_{i}-\hat{y}_{i}}{y_{i}}\right)^{2}}
        \end{equation}
        RMRSE is also sometimes reported as a percentage, which is the above equation multiplied by 100.
    \item \textbf{RRMSE}. Similarly to WAPE, it takes the total squared error and normalizes it by dividing by the total squared error of a simple predictor. By taking the square root of the relative squared error one reduces the error to the same dimensions as the quantity being predicted. In formula: 
        \begin{equation}
            RRMSE \coloneqq \sqrt{ \frac{\sum_{i=1}^{n} \left(y_{i}-\hat{y}_{i}\right)^{2}}{\sum_{i=1}^{n} \left(y_{i}\right)^{2}}}
        \end{equation}
        RRMSE is also sometimes reported as a percentage, which is the above equation multiplied by 100.
\end{itemize}
Both MAE and RMSE express average model prediction error in units of the variable of interest, they can range from 0 to $\infty$ and are indifferent to the direction of errors. They are negatively-oriented scores, which means lower values are better. Since the errors are squared before they are averaged, the RMSE gives a relatively high weight to large errors. This means the RMSE should be more useful when large errors are particularly undesirable and generally, RMSE will be higher than or equal to MAE. It can be noted that RMSRE and RRMSE are completely analogous to MAPE and WAPE where the absolute value is replaced with the square. RMSRE and MAPE are the relative versions of RMSE and MAE respectively and are taken into consideration in this context because, for example, an error of \EUR{100} out of \EUR{200} is worse than an error of the same amount out of \EUR{2000}. However, they have some drawbacks: they are undefined for data points where the target value is $0$ and it can grow unexpectedly large if the actual values are exceptionally small themselves. To avoid these problems, an arbitrarily small term is usually added to the denominator. Moreover, they are asymmetric and it puts a heavier penalty on negative errors (when forecasts are higher than target) than on positive errors. To solve these problems RRMSE and WAPE are introduced and they are particularly recommended when the number of samples is low or their values are on different scales. 

\newpage
\section{G1++ parameters}
We present the values of the G1 ++ parameters used for the creation of the dataset
\label{App:hw}
\begin{longtable}{Xcc}
\hline
\textbf{Speed of mean reversion ($a$)} & \textbf{Volatility ($\sigma$)} \\
\hline
-2\% & 0.5\% \\
-1\% & 1\% \\
2\% & 5\% \\
3\% & 2\% \\
4\% & 1.5\% \\
5\% & 2.5\% \\
6\% & 3\% \\
9\% & 4\% \\
15\% & 7\% \\
30\% & 8\% \\
\hline
\caption{G1++ parameters choosen for the creation of the dataset.}
\label{Tab:hw}
\end{longtable}

\newpage
\section{Market data}
\label{App:market}

\begin{longtable}{Xcccc}
\hline
\textbf{Maturity} & \textbf{ZC rates (EONIA OIS)} & \textbf{ZC rates EURIBOR 6M} \\
\hline
01/11/19&	-0.46943&   	-0.32479\\
04/11/19&	-0.46944&  	    -0.32479\\   
05/11/19&	-0.46911& 	    -0.32479\\   
11/11/19&	-0.46299&   	-0.32479\\   
18/11/19&	-0.46237&   	-0.32479\\  
25/11/19&	-0.46097&   	-0.32479\\   
04/12/19&	-0.46056&   	-0.32479\\   
06/01/20&	-0.46293&   	-0.32808\\   
04/02/20&	-0.46483&   	-0.33032\\   
04/03/20&	-0.46782&   	-0.33337\\   
06/04/20&	-0.47286&   	-0.33823\\   
04/05/20&	-0.47693&   	-0.34200\\   
04/06/20&	-0.48104&   	-0.34571\\   
06/07/20&	-0.48517&   	-0.34914 \\  
04/08/20&	-0.48830&   	-0.35194 \\  
04/09/20&	-0.49245&   	-0.35463 \\  
05/10/20&	-0.49560&   	-0.35705  \\ 
04/11/20&	-0.49875&   	-0.35916  \\ 
04/12/20&	-0.50170&   	-0.36108  \\ 
04/01/21&	-0.50438&   	-0.36283  \\ 
04/02/21&	-0.50669&   	-0.36432  \\
04/03/21&	-0.50847&   	-0.36541   \\
06/04/21&	-0.51032&   	-0.36636   \\
04/05/21&	-0.51175&   	-0.36686   \\
04/06/21&	-0.51321&   	-0.36706   \\
05/07/21&	-0.51435&   	-0.36692   \\
04/08/21&	-0.51490&   	-0.36648   \\
06/09/21&	-0.51467&   	-0.36571   \\
04/10/21&	-0.51402&   	-0.36484   \\
04/11/21&	-0.51305&   	-0.36368   \\
04/11/22&	-0.50001&   	-0.33954   \\
06/11/23&	-0.47279&   	-0.30385   \\
04/11/24&	-0.43445&   	-0.26003   \\
04/11/25&	-0.38600&   	-0.20939   \\
04/11/26&	-0.33043&   	-0.15369   \\
04/11/27&	-0.27072&   	-0.09292   \\
06/11/28&	-0.20884&   	-0.03198   \\
05/11/29&	-0.14779&   	 0.02709   \\
04/11/30&	-0.08655&   	 0.08475   \\
04/11/31&	-0.03021&   	 0.13987   \\
04/11/32&	 0.02316&   	 0.19116   \\
04/11/33&	 0.07304&   	 0.23762   \\
06/11/34&	 0.11844&   	 0.27911   \\
05/11/35&	 0.15768&   	 0.31473   \\
04/11/36&	 0.19140&   	 0.34514   \\
04/11/37&	 0.21996&   	 0.37064   \\
04/11/38&	 0.24384&   	 0.39166   \\
04/11/39&	 0.26351&   	 0.40860   \\
05/11/40&	 0.27957&   	 0.42199   \\
04/11/41&	 0.29222&   	 0.43210   \\
04/11/42&	 0.30202&   	 0.43944   \\
04/11/43&	 0.30928&   	 0.44437   \\
04/11/44&	 0.31438&   	 0.44724   \\
06/11/45&	 0.31783&   	 0.44865   \\
05/11/46&	 0.31991&   	 0.44894   \\
04/11/47&	 0.32080&   	 0.44821   \\
04/11/48&	 0.32064&   	 0.44655   \\
04/11/49&	 0.31953&   	 0.44403   \\
04/11/54&	 0.30413&   	 0.42202   \\
04/11/59&	 0.27804&   	 0.38979   \\
04/11/69&	 0.22309&   	 0.32341   \\
06/11/79&	 0.19565&   	 0.28838   \\
\hline
\caption{EONIA OIS (second column) and EURIBOR 6M (third column) yield curve as of 31st October 2019. Discount rates are zero coupon rates expressed in percentage values, continuous compounding, act/365 daycount convention.}
\label{Tab:ois}
\end{longtable}

\begin{table}[H]
\centering
\scalebox{0.9}{
\begin{tabular}{ccccccccccccccc}
\toprule
 & 1Y & 2Y & 3Y & 4Y & 5Y & 6Y & 7Y & 8Y & 9Y & 10Y & 15Y & 20Y & 25Y & 30Y\\
\midrule
1M & -0.44 & -0.36 & -0.34 & -0.30 & -0.26 & -0.20 & -0.15 & -0.09 & -0.03 & 0.03 & 0.28 & 0.40 & 0.44 & 0.44 \\
2M &-0.44 & -0.36 & -0.34 & -0.30 & -0.25 & -0.20 & -0.14 & -0.08 & -0.02 & 0.04 & 0.29 & 0.41 & 0.44 & 0.44 \\
3M &-0.45 & -0.36 & -0.33 & -0.29 & -0.24 & -0.19 & -0.13 & -0.07 & -0.01 & 0.05 & 0.29 & 0.41 & 0.45 & 0.44 \\
6M &-0.45 & -0.36 & -0.32 & -0.28 & -0.23 & -0.17 & -0.11 & -0.05 & 0.01 & 0.07 & 0.31 & 0.43 & 0.46 & 0.45 \\
9M &-0.45 & -0.35 & -0.30 & -0.26 & -0.20 & -0.15 & -0.08 & -0.02 & 0.04 & 0.10 & 0.33 & 0.44 & 0.46 & 0.45 \\
1Y &-0.44 & -0.33 & -0.29 & -0.24 & -0.18 & -0.12 & -0.06 & 0.01 & 0.07 & 0.13 & 0.35 & 0.45 & 0.47 & 0.46 \\
18M &-0.41 & -0.29 & -0.24 & -0.19 & -0.13 & -0.07 & 0.00 & 0.06 & 0.12 & 0.18 & 0.39 & 0.48 & 0.49 & 0.47 \\
2Y &-0.37 & -0.25 & -0.19 & -0.14 & -0.07 & -0.01 & 0.06 & 0.12 & 0.18 & 0.24 & 0.43 & 0.50 & 0.51 & 0.49 \\
3Y &-0.27 & -0.14 & -0.08 & -0.02 & 0.05 & 0.12 & 0.18 & 0.24 & 0.29 & 0.34 & 0.50 & 0.55 & 0.54 & 0.51 \\
4Y &-0.16 & -0.02 & 0.04 & 0.11 & 0.18 & 0.24 & 0.30 & 0.36 & 0.40 & 0.45 & 0.57 & 0.59 & 0.56 & 0.52 \\
5Y &-0.03 & 0.11 & 0.18 & 0.25 & 0.31 & 0.37 & 0.42 & 0.45 & 0.51 & 0.54 & 0.63 & 0.62 & 0.58 & 0.54 \\
7Y & 0.26 & 0.38 & 0.44 & 0.50 & 0.55 & 0.59 & 0.62 & 0.65 & 0.67 & 0.69 & 0.70 & 0.66 & 0.60 & 0.55 \\
10Y &0.64 & 0.70 & 0.74 & 0.76 & 0.78 & 0.79 & 0.80 & 0.78 & 0.80 & 0.79 & 0.73 & 0.66 & 0.59 & 0.52 \\
15Y &0.85 & 0.84 & 0.83 & 0.82 & 0.80 & 0.78 & 0.76 & 0.74 & 0.72 & 0.70 & 0.61 & 0.53 & 0.46 & 0.40 \\
20Y & 0.68 & 0.67 & 0.64 & 0.66 & 0.60 & 0.58 & 0.56 & 0.55 & 0.53 & 0.51 & 0.44 & 0.37 & 0.32 & 0.27 \\
25Y & 0.49 & 0.47 & 0.46 & 0.44 & 0.43 & 0.41 & 0.34 & 0.39 & 0.37 & 0.36 & 0.29 & 0.24 & 0.20 & 0.17 \\
30Y & 0.33 & 0.33 & 0.32 & 0.30 & 0.29 & 0.27 & 0.26 & 0.24 & 0.23 & 0.22 & 0.18 & 0.14 & 0.12 & 0.13 \\
\bottomrule
\end{tabular}}
\caption{EUR ATM Swaption forward rates as of 31st October 2019. Columns: Swaption tenors; rows: Swaption expiries.}
\label{Tab:forward}
\end{table}

\begin{table}[H]
\centering
\scalebox{0.9}{
\begin{tabular}{ccccccccccccccc}
\toprule
 & 1Y & 2Y & 3Y & 4Y & 5Y & 6Y & 7Y & 8Y & 9Y & 10Y & 15Y & 20Y & 25Y & 30Y\\
\midrule
1M &3.5 & 9 & 16 & 25.5 & 36 & 47.5 & 60.5 & 74 & 88 & 103 & 175 & 246 & 307 & 369 \\
2M & 5.5 & 13 & 23.5 & 36.5 & 49 & 67 & 87.5 & 108 & 129 & 150 & 253 & 357 & 449 & 539 \\
3M & 6.5 & 15.5 & 28 & 45.5 & 62.5 & 84 & 109 & 134 & 160 & 189 & 316 & 445 & 552 & 672 \\
6M & 9 & 22 & 39 & 62 & 87.5 & 116 & 149 & 183 & 220 & 257 & 430 & 596 & 750 & 898 \\
9M & 11.5 & 27.5 & 50.5 & 79 & 109 & 144 & 182 & 222 & 265 & 311 & 515 & 709 & 889 & 1066 \\
1Y & 14.5 & 33.5 & 60.5 & 93 & 128 & 168 & 212 & 259 & 308 & 359 & 582 & 802 & 1007 & 1211 \\
18M & 20 & 46 & 81 & 121 & 165 & 216 & 270 & 324 & 383 & 445 & 706 & 965 & 1203 & 1459 \\
2Y & 26.5 & 61 & 104 & 152 & 204 & 261 & 323 & 386 & 453 & 524 & 819 & 1105 & 1379 & 1662 \\
3Y & 43 & 92.5 & 150 & 213 & 282 & 351 & 429 & 501 & 580 & 661 & 1002 & 1337 & 1662 & 1982 \\
4Y & 59 & 123 & 196 & 272 & 352 & 433 & 518 & 606 & 694 & 779 & 1159 & 1528 & 1878 & 2228 \\
5Y & 75.5 & 155 & 238 & 327 & 419 & 510 & 603 & 698 & 794 & 888 & 1297 & 1699 & 2073 & 2455 \\
7Y & 105.5 & 211 & 319 & 427 & 535 & 644 & 752 & 859 & 969 & 1079 & 1549 & 2004 & 2416 & 2835 \\
10Y & 140.5 & 279 & 416 & 549 & 679 & 810 & 939 & 1064 & 1188 & 1322 & 1880 & 2404 & 2887 & 3371 \\
15Y & 173.5 & 343 & 511 & 673 & 832 & 990 & 1141 & 1289 & 1444 & 1595 & 2247 & 2858 & 3407 & 3946 \\
20Y & 198 & 392 & 585 & 766 & 949 & 1133 & 1304 & 1476 & 1643 & 1809 & 2552 & 3213 & 3806 & 4377 \\
25Y & 216 & 429 & 640 & 843 & 1041 & 1237 & 1422 & 1602 & 1787 & 1966 & 2787 & 3491 & 4118 & 4711 \\
30Y & 229.5 & 458 & 683 & 901 & 1118 & 1325 & 1519 & 1702 & 1882 & 2067 & 2930 & 3677 & 4335 & 4938 \\
\bottomrule
\end{tabular}}
\caption{EUR ATM European Swaption straddles, forward premium, physical LCH settlment, EONIA discounting, notional 10,000 \euro. Date and table format as in Tab. \ref{Tab:forward}.}
\label{Tab:straddle}
\end{table}

\newpage
\section{Bermudan basket}
\label{App:basket}

\begin{figure}[H]
\centering
\includegraphics[width=1\linewidth]{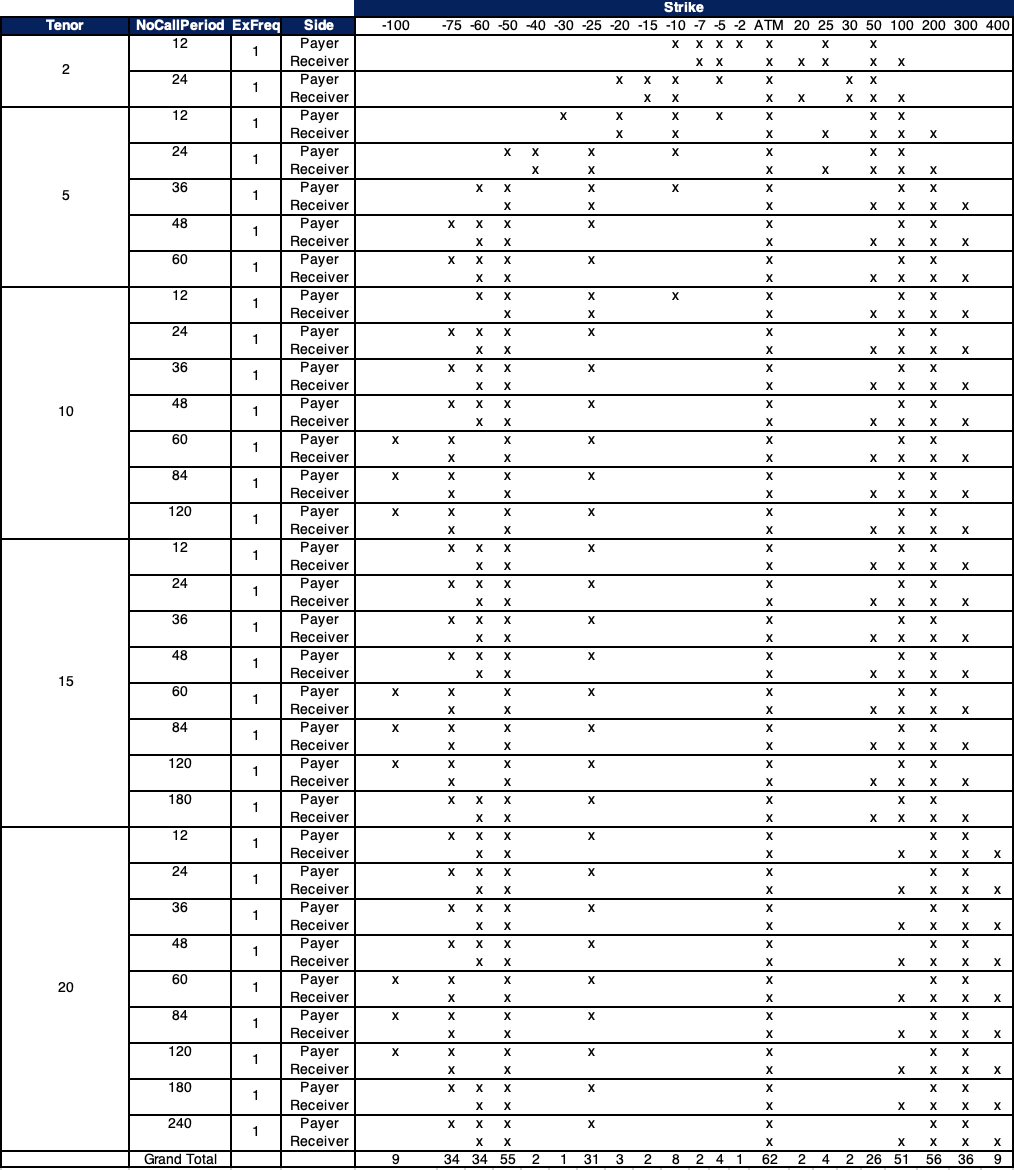}
\caption{Bermudan Swaption selected for the dataset.}
\label{Fig:basket}
\end{figure}

\end{appendices}
\newpage
%--- Bibliography
\bibliographystyle{unsrt}
\bibliography{FinanceBibliography}

\end{document}

%% file: Newcommands.tex
% --- Definitions of new Latex commands ---
%
% REMIND: always keep a single file with all definitions, beware of multiple copies.

% --- Text ---
\newcommand{\ie}{i.e.~} 	                    % i.e. with non-breakable space
\newcommand{\eg}{e.g.~} 	                    % e.g. with non-breakable space
\newcommand{\rhs}{r.h.s.~} 	                    % r.h.s. with non-breakable space
\newcommand{\wrt}{w.r.t.~} 	                    % w.r.t. with non-breakable space
\newcommand{\old}[1]{\textcolor{blue}{#1}}      % old text, blue, strike-through tex
\newcommand{\new}[1]{\textcolor{red}{#1}}       % new text, red, strike-through tex
\newcommand{\textbox}[1]{\mbox{\textit{#1}}} 	% text inside math
\newcommand{\virgolette}[1]{``#1''}

% --- Mathematical Definitions ---
\newtheorem{theorem}{Theorem}[section]
\newtheorem{definition}[theorem]{Definition}
\newtheorem{definitions}[theorem]{Definition}
\newtheorem{proposition}[theorem]{Proposition}
\newtheorem{remark}[theorem]{Remark}
\newtheorem{corollary}[theorem]{Corollary}
\newtheorem{example}[theorem]{Example}
\newtheorem{lemma}[theorem]{Lemma}

% --- Environments ---
\newenvironment{system}{\left\lbrace\begin{array}{@{}l@{}}}{\end{array}\right.}

% --- Delimiters ---
\newcommand{\Parenthesis}[1]{\left( #1 \right)}         % (xxx)
\newcommand{\Brack}[1]{\left\lbrack #1 \right\rbrack}   % [xxx]
\newcommand{\Brace}[1]{\left\lbrace #1 \right\rbrace}   % {xxx}
\newcommand{\Abs}[1]{\left\lvert #1 \right\rvert}       % |xxx|
\newcommand{\Norm}[1]{\left\lVert #1 \right\rVert}      % ||xxx||
\newcommand{\Mean}[1]{\left\langle #1 \right\rangle}    % <xxx>
\newcommand{\QuoteDouble}[1]{``#1''}                    % double quotation marks
\newcommand{\QuoteSingle}[1]{`#1'}                      % single quotation marks

% --- Math operators ---
\newcommand{\beq}{\begin{equation}}         % \begin{equation}
\newcommand{\eeq}{\end{equation}}           % \end{equation}
\newcommand{\Max}{\mbox{Max}}                                   % Max operator
\newcommand{\Min}{\mbox{Min}}                                   % Min operator
\newcommand{\Derp}[2]{\frac{\partial #1}{\partial #2}}          % partial derivative, 1st order
\newcommand{\DerpXX}[2]{\frac{\partial^2 #1}{\partial #2 ^2}}   % partial derivative, 2nd order
\newcommand{\DerpXY}[3]{\frac{\partial^2 #1}{\partial #2 \partial #3}}    % partial derivative, 2nd order, cross
\newcommand{\half}[0]{\frac{1}{2}}                                                  % one half
\newcommand{\ind}[1]{\mathbf{1}_{#1}} 	                                            % indicator
\newcommand{\E}[1]{\mathbb{E}\left[#1\right]}                                       % Expected value
\newcommand{\expt}[1]{\mathbb{E}_t\left[#1\right]}                                  % Expected value at time t
\newcommand{\Econd}[2]{\mathbb{E}\left[#1\;\middle \vert\;\mathcal{F}_{#2}\right]}  % Expected value conditioned

% --- Sets ---
\newcommand{\Nset}{\mathbb{N}}      % Natural numbers set
\newcommand{\Zset}{\mathbb{Z}}      % Integer numbers set
\newcommand{\Qset}{\mathbb{Q}}      % Rational numbers set
\newcommand{\Rset}{\mathbb{R}}      % Real numbers set

% --- Finance ---
\newcommand{\Black}{\mbox{Black}}				    % Black's formula
\newcommand{\XVA}{\mbox{\textit{XVA}}}				% XVA
\newcommand{\CVA}{\mbox{\textit{CVA}}}				% CVA
\newcommand{\DVA}{\mbox{\textit{DVA}}}				% DVA
\newcommand{\FVA}{\mbox{\textit{FVA}}}				% FVA

% --- Financial Instruments ---
\newcommand{\Depo}{\mbox{\textbf{Depo}}}			% Deposit
\newcommand{\FRA}{\mbox{\textbf{FRA}}}			    % Forward Rate Agreement
\newcommand{\Futures}{\mbox{\textbf{Futures}}}      % Futures
\newcommand{\ZCB}{\mbox{ZCB}}					    % Zero Coupon Bond
\newcommand{\Swap}{\mbox{\textbf{Swap}}}			% Swap
\newcommand{\Swaplet}{\mbox{\textbf{Swaplet}}}		% Swaplet
\newcommand{\IRS}{\mbox{\textbf{IRS}}}			    % Interest Rate Swap
\newcommand{\IRSlet}{\mbox{\textbf{IRSlet}}}        % Interest Rate Swaplet
\newcommand{\OIS}{\mbox{\textbf{OIS}}}              % OIS (Overnight Indexed Swap)
\newcommand{\OISlet}{\mbox{\textbf{OISlet}}}        % OISlet
\newcommand{\BSwap}{\mbox{\textbf{BSwap}}}          % Basis Swap
\newcommand{\BSwaplet}{\mbox{\textbf{BSwaplet}}}    % Basis Swaplet
\newcommand{\IRBS}{\mbox{\textbf{IRBS}}}            % Interest Rate Basis Swap
\newcommand{\IRBSlet}{\mbox{\textbf{IRBSlet}}}      % Interest Rate Basis Swaplet
\newcommand{\CCSwap}{\mbox{\textbf{CCSwap}}}        % Cross Currency Swap
\newcommand{\CCS}{\mbox{\textbf{CCS}}}              % Cross Currency Swap
\newcommand{\CCSwaplet}{\mbox{\textbf{CCSwaplet}}}  % Cross Currency Swaplet
\newcommand{\CCSlet}{\mbox{\textbf{CCSlet}}}        % Cross Currency Swaplet
\newcommand{\Caplet}{\mbox{\textbf{Caplet}}}        % Caplet
\newcommand{\Floorlet}{\mbox{\textbf{Floorlet}}}    % Floorlet
\newcommand{\cf}{\mbox{\textbf{cf}}}                % Caplet/Floorlet
\newcommand{\CAP}{\mbox{\textbf{Cap}}}              % Cap
\newcommand{\Floor}{\mbox{\textbf{Floor}}}          % Floor
\newcommand{\CF}{\mbox{\textbf{CF}}}                % Cap/Floor
\newcommand{\Swaption}{\mbox{\textbf{Swaption}}}    % Swaption
\newcommand{\CMSlet}{\mbox{\textbf{CMSlet}}}        % CMSlet
\newcommand{\CMS}{\mbox{\textbf{CMS}}}              % CMS
\newcommand{\CMScf}{\mbox{\textbf{CMScf}}}          % CMS caplet/floorlet
\newcommand{\FXFwd}{\mbox{\textbf{FXFwd}}}          % Forex forward

%% file: paper.bbl
\begin{thebibliography}{10}

\bibitem{Gla03}
Paul Glasserman.
\newblock {\em Monte {C}arlo Methods in Financial Engineering}.
\newblock Springer, 2003.

\bibitem{Bec20}
Sebastian Becker, Patrick Cheridito, and Arnulf Jentzen.
\newblock Deep optimal stopping, 2020.

\bibitem{Gas20}
Raquel~M. Gaspar, Sara~D. Lopes, and Bernardo Sequeira.
\newblock Neural network pricing of american put options.
\newblock {\em Risks}, 8(3), 2020.

\bibitem{BecCher20}
Sebastian Becker, Patrick Cheridito, and Arnulf Jentzen.
\newblock Pricing and hedging american-style options with deep learning.
\newblock {\em Journal of Risk and Financial Management}, 13:158, 07 2020.

\bibitem{Lap20}
Bernard Lapeyre and Jérôme Lelong.
\newblock Neural network regression for bermudan option pricing, 2020.

\bibitem{Gou19}
Ludovic Goudenège, Andrea Molent, and Antonino Zanette.
\newblock Variance reduction applied to machine learning for pricing
  bermudan/american options in high dimension, 2019.

\bibitem{HulWhi1994}
John Hull and Alan White.
\newblock Numerical procedures for implementing term structure models {I}:
  Single-factor models.
\newblock {\em Journal of Derivatives}, 2:7--16, 1994.

\bibitem{LonSch1998}
Francis Longstaff and Eduardo Schwartz.
\newblock Valuing american options by simulation: A simple least-squares
  approach.
\newblock Working paper, The Anderson School, UCLA, 1998.

\bibitem{Hag02}
Patrick Hagan.
\newblock In thetrenches adjusters: Turning good prices into great prices.
\newblock {\em Wilmott}, 2002:56--59, 12 2002.

\bibitem{has01}
Trevor Hastie, Robert Tibshirani, and Jerome Friedman.
\newblock {\em The Elements of Statistical Learning}.
\newblock Springer Series in Statistics. Springer New York Inc., New York, NY,
  USA, 2001.

\bibitem{ger17}
Aurélien Géron.
\newblock {\em Hands-on machine learning with Scikit-Learn and TensorFlow :
  concepts, tools, and techniques to build intelligent systems}.
\newblock O'Reilly Media, Sebastopol, CA, 2017.

\bibitem{Bre01}
L.~Breiman.
\newblock Random forest.
\newblock {\em Machine Learning}, 45:5 -- 32, 2001.

\bibitem{Fri02}
Jerome~H. Friedman.
\newblock Stochastic gradient boosting.
\newblock {\em Computational Statistics \& Data Analysis}, 38(4):367--378,
  2002.

\end{thebibliography}
